\documentclass[%
 reprint,
superscriptaddress,
 amsmath,amssymb,
 aps,
]{revtex4-2}

\usepackage{graphicx}% Include figure files
\usepackage{dcolumn}% Align table columns on decimal point
\usepackage{bm}% bold math
\begin{document}

\preprint{APS/123-QED}

\title{Wood compression in 4D in-situ tomography}% Force line breaks with \\

\author{Tero M\"{a}kinen}
 \email{Corresponding author\\tero.j.makinen@aalto.fi}
\affiliation{%
 Department of Applied Physics, Aalto University,
 P.O.  Box 11100, 00076 Aalto, Espoo, Finland
}%
\affiliation{NOMATEN Centre of Excellence, National Centre for Nuclear Research, ul. A. Soltana 7, 05-400  Otwock-\'{S}wierk, Poland}
\author{Alisa Halonen}
\affiliation{%
 Department of Applied Physics, Aalto University,
 P.O.  Box 11100, 00076 Aalto, Espoo, Finland
}%
\author{Juha Koivisto}%
\affiliation{%
 Department of Applied Physics, Aalto University,
 P.O.  Box 11100, 00076 Aalto, Espoo, Finland
}%

\author{Mikko J. Alava}%
\affiliation{%
 Department of Applied Physics, Aalto University,
 P.O.  Box 11100, 00076 Aalto, Espoo, Finland
}%
\affiliation{NOMATEN Centre of Excellence, National Centre for Nuclear Research, ul. A. Soltana 7, 05-400  Otwock-\'{S}wierk, Poland}

\date{\today}% It is always \today, today,

\begin{abstract}
Wood deformation, in particular when subject to compression exhibits scale-free avalanche like behaviour as well as structure dependent localization of deformation. We have taken 3D x-ray tomographs during compression with constant stress rate loading. Using Digital Volume Correlation we obtain the local total strain during the experiment and compare it to the global strain and acoustic emission.
The wood cells collapse layer by layer throughout the sample starting from the softest parts, the spring wood. 
As the damage progresses, more and more of the softwood layers throughout the sample collapse which indicates damage spreading instead of localization.
In 3D one can see a fat-tailed local strain rate distribution indicating that inside the softwood layers the damage occurs in localized spots. The observed log-normal strain distribution is in agreement with this view of the development of independent local collapses or irreversible deformation events.
A key feature in the mechanical behavior of wood is then in the complex interaction of localized deformation between or among the annual rings. 
\end{abstract}

\maketitle

%\tableofcontents

\section{Introduction}

Damage localization in materials under compressive loading leads to unpredictable behavior, which has in recent years an active area of research in theoretical \cite{berthier2017damage,berthier2021damage,mayya2022criticality}, numerical~\cite{castellanos2018avalanche} as well as experimental studies on e.g. rocks~\cite{lockner1993role, davidsen2007scaling},
porous materials~\cite{baro2013statistical}, solid foams~\cite{makinen2020crossover, reichler2021scalable}, and wood~\cite{makinen2015avalanches}.
Compared to tensile failure, compression is particularly interesting as structures can carry load even after deforming \cite{keckes2003cell} through frictional contacts.
Here, we take an ordinary wood sample as an example of heterogeneous cellular material \cite{gibson_ashby_1997} and crush it under constant stress rate loading while recording the 3D deformation using fast synchrotron tomography.\\

X-ray tomography \cite{maire2014quantitative, withers2021x} has been growing in popularity as a research tool for material science. It is a method for acquiring 3D images of samples and in in-situ setups can provide sequences of 3D images during an experiment (4D imaging).
It has been extensively used e.g. to observe faulting in rock fracture \cite{renard2017microscale, renard2018critical, kandula2019dynamics, cartwright2020catastrophic, kandula2021synchrotron} where the focus is on determining the locations of microcracks, which can be done directly from the reconstructed volumes by image segmentation \cite{iassonov2009segmentation}.\\

Used in conjunction with tomography, Digital Volume Correlation (DVC) \cite{buljac2018digital} is a powerful method for determining the local strains in 3D from a set of tomography images. It has been used to study the deformation of a wide variety of different materials, such as metals \cite{li2020situ}, composites \cite{mendoza2019differentiating, holmes2022digital}, bone \cite{fernandez2018effect, fernandez2021time}, and coal~\cite{vishal2022mechanical}.
The advances in computational capabilities now enable the DVC determination of strain fields even for a large number of tomography images, which with fast tomography leads to a good time-resolution of the 4D imaging.\\

Our test material wood is an ubiquitous biological material with a cellular structure. The arrangement of wood cells depends on the annual growth cycle, which leads to a complex hierarchical structure of alternating softwood and hardwood layers \cite{ando1999mechanism, stanzl2011wood}. This macroscopic structure -- annual rings -- and especially its orientation \cite{miksic2013effect} has a large effect on the mechanical properties of wood.\\

In our previous work \cite{makinen2015avalanches} we studied the avalanche behavior in wood compression using Digital Image Correlation (DIC) and acoustic emission (AE). Scale-free avalanche behavior, or crackling noise \cite{sethna2001crackling}, was observed in terms of the AE energies reminiscent of compression of porous brittle media \cite{baro2013statistical, xu2019criticality} and earthquakes \cite{utsu1999representation}.
However concomitantly we also observed structure-dependent strain localization (collapse of softwood layers) which somewhat contradicts this scalefree picture.\\

The aim of this paper is to explore the time-evolution of this structure-dependent localization in 3D.
The main goal is to determine, if imaging the whole bulk sample instead of just one surface yields additional information and if this information can be used to explain the disconnect between structure-dependency and scalefree behavior.

\section{Methods} 

\begin{figure}[t!]
\includegraphics[width=\columnwidth]{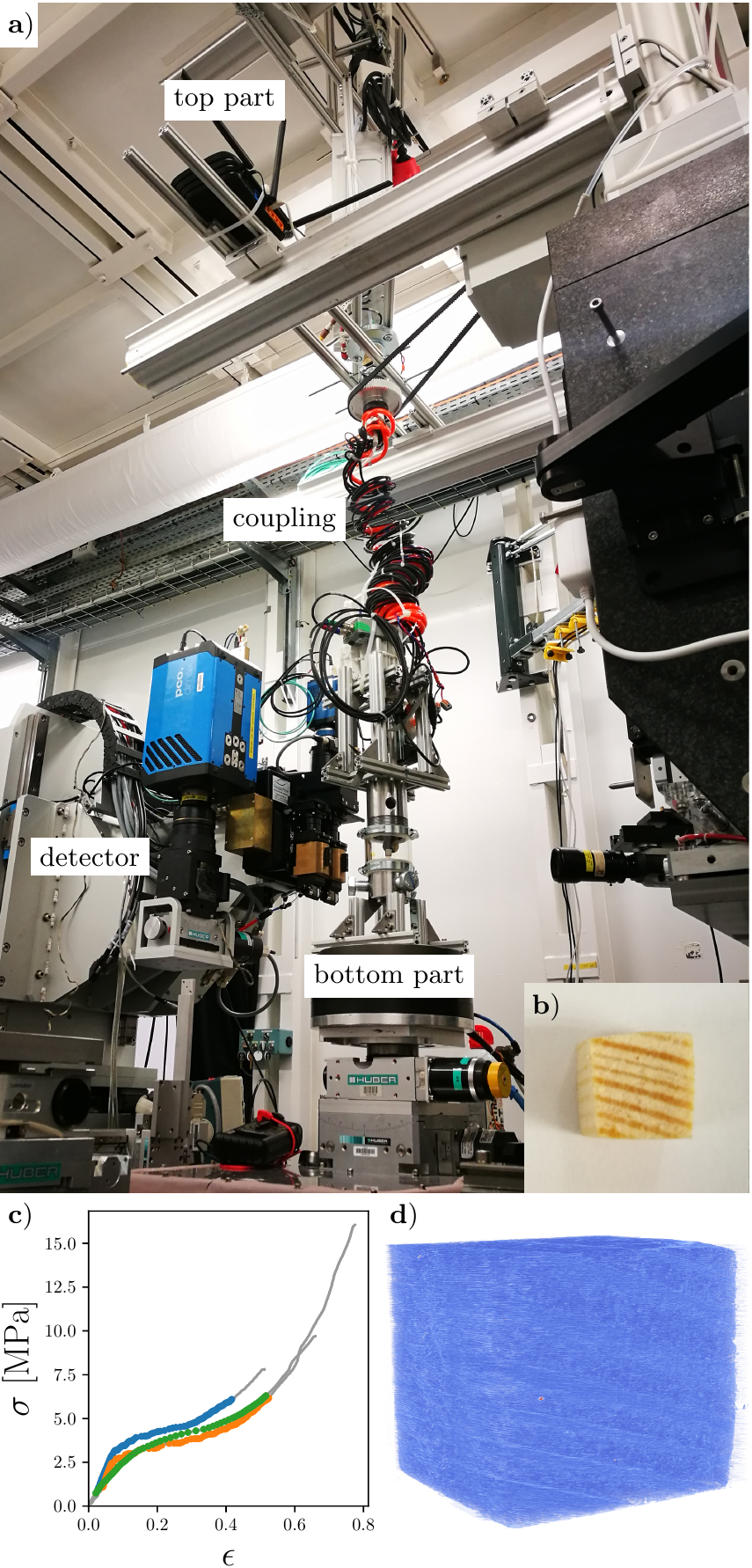}
\caption{\label{fig:setup} \textbf{a}) The experimental setup showing the bottom part on the rotation stage of the beamline and the top part above as well as the coupling of the two parts.
\textbf{b}) An image of a sample used, showing a typical annual ring structure.
\textbf{c}) The measured stress-strain curves (gray) and the imaged portion of the compression (colored) for the three experiments considered. The following figures focus on the experiment plotted in blue.
\textbf{d}) The reconstructed sample volume.}
\end{figure}

The experiments were performed on the ID15A beamline at the European Synchrotron Radiation Facility (ESRF). The experimental setup consists of a compression device similar to the Mj\"olnir \cite{butler2020mjolnir} and the whole setup is shown in Fig.~\ref{fig:setup}a. Our compression device consists of two individually rotating parts coupled together via cables. The bottom part lies on the rotation stage of the beamline and houses a rotationally symmetric sample holder, a compression piston, a load sensor and a displacement sensor. The top part is rotated separately from the bottom part using a stepper motor and it houses the data acquisition units, power supply and a WiFi connection to the control hutch.\\

The samples (seen in Fig.~\ref{fig:setup}b) were $6.2 \pm 0.3$ mm $\times$ $6.2 \pm 0.1$ mm $\times$ $5.9 \pm 0.2$ mm cubes (in order height, width, depth and compressed from the top) of dry pine (\emph{Pinus sylvestris}) which means that in a sample one has 4-6
annual rings. The compression is applied in direction perpendicular to this annual ring orientation, although there is a slight angle between the compression direction and the annual ring orientation.\\

The samples were compressed using a constant stress rate corresponding to a force rate of 3 N/s, which results in a stress rate of $82 \pm 3$ kPa/s.
We record the global deformation of the sample using the displacement sensor (giving the engineering strain $\epsilon = d / h$ where $d$ is the displacement and $h$ the initial sample height) and the load sensor (giving the engineering stress $\sigma = F / A$ where $F$ is the applied force and $A$ the initial cross-sectional area of the sample). This gives us a time-series for the strain $\epsilon$ and stress $\sigma$ recorded at a frequency of 102.4~kHz (see Fig.~\ref{fig:setup}c). After an application of a moving average, a strain rate $\dot{\epsilon}$ was calculated from the strain time-series using numerical differentiation.\\

Due to the large amount of data produced only part of the tomography data for the whole compression was recorded, namely part of the initial elastic regime and the final densification regimes were discarded.
The camera of the beamline operates with an acquisition frequency of 500 Hz and the rotation stage performes half a rotation per second, leading to a time-resolution of one full tomography image per second and an angular resolution of 0.36 degrees per projection.
The imaging resolution (voxel size) is 11 $\mu$m and the region of interest in the DVC computations is a cube with side length of 10 voxels (110 $\mu$m). These regions of interest are placed 6 voxels apart, giving a DVC resolution of 66 $\mu$m.

\begin{figure*}[ht!]
    \centering
    \includegraphics[width=\textwidth]{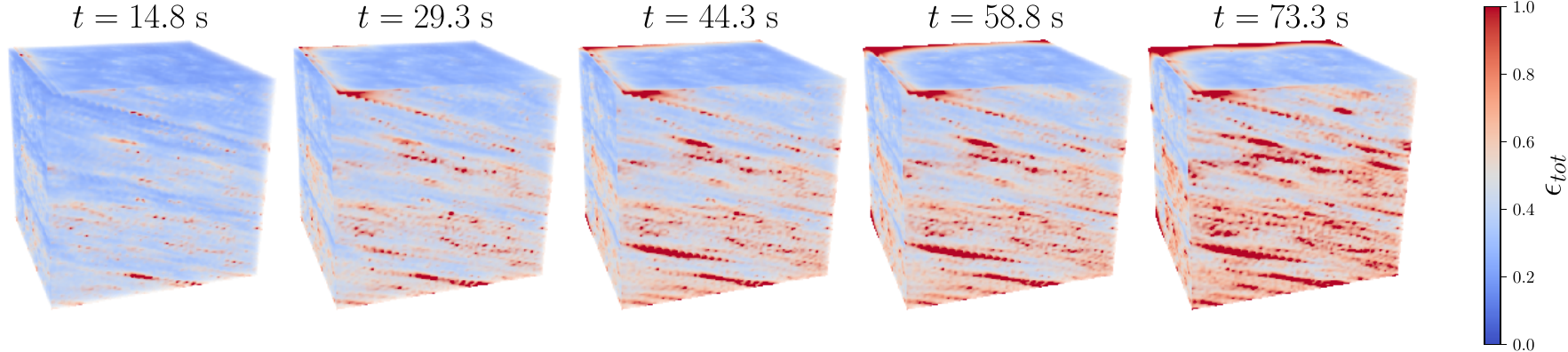}%
    \caption{The time evolution of the strain in the wood sample. The color code indicates the local total strain. The damage starts from hotspots in the softwood layers and then spreads throughout the layer and to the other softwood layers.
    }
    \label{fig:imgSet}
\end{figure*}

\begin{figure}[t!]
    \centering
    \includegraphics[width=\columnwidth]{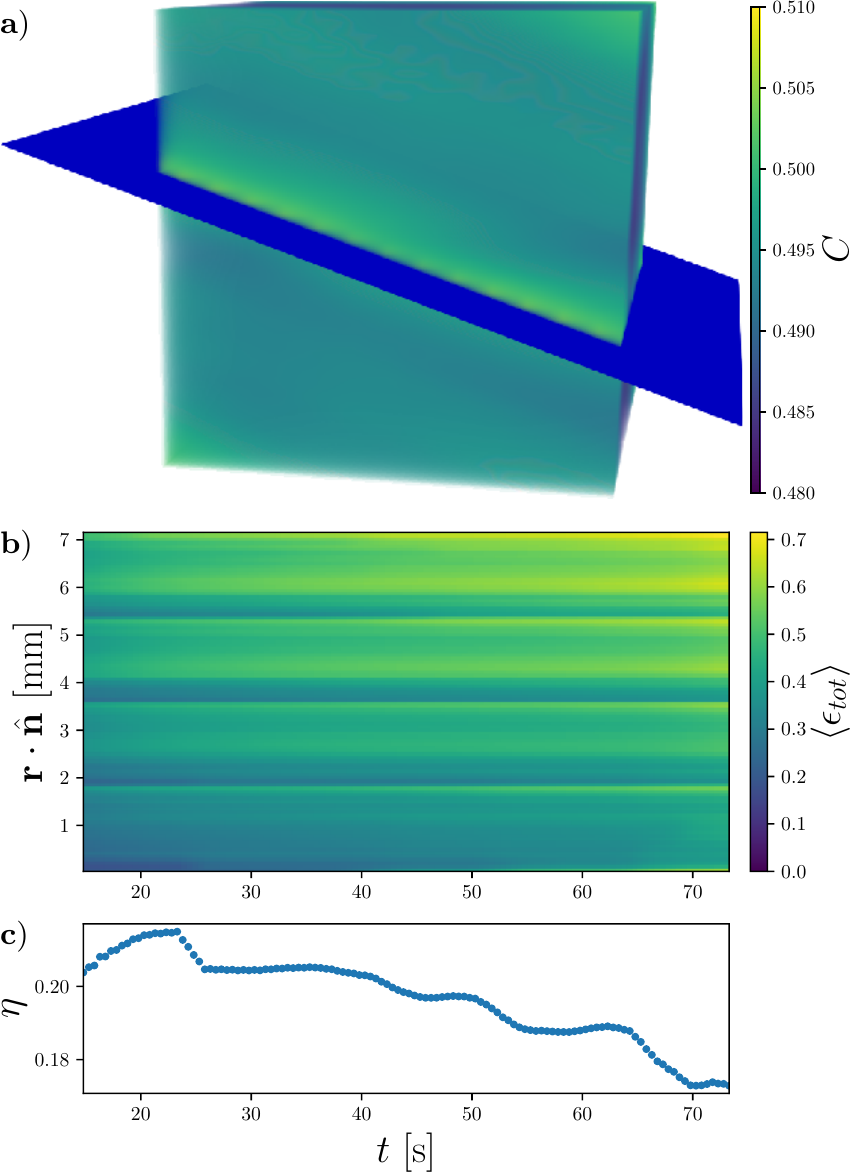}
    \caption{\textbf{a}) The autocorrelation function $C$ (Eq.~\ref{eq:autocorrelation}) of the local total strain field computed in a 0.44~mm $\times$ 0.44~mm $\times$ 0.44~mm box, showing the alignment of the correlations with the orientation of the annual rings. The blue plane is fitted to this annual ring orientation (direction of high correlations) and is used for the dimensionality reduction in panel b.
    \textbf{b}) A spacetime plot of the average local strain $\langle \epsilon_{tot} \rangle$ in the direction perpendicular to the fitted plane (see panel a), described by the projection of the position vector $\bm{r}$ onto the unit normal of the plane $\bm{\hat{n}}$. The time value $t=0$ s corresponds to the start of the test and tomography images are collected only after approximately 15 s.
    \textbf{c}) The localization measure $\eta$ as a function of time $t$ (time axis aligned with panel b) showing the delocalization of strain.
    }
    \label{fig:projection}
\end{figure}

The reconstructions of the 3D volumes from the projection data were done using the simultaneous iterative reconstruction technique (SIRT) algorithm implemented in the ASTRA Toolbox \cite{van2016fast, van2015astra, palenstijn2011performance}. The reconstructions (an example can be seen in Fig.~\ref{fig:setup}d) were computed from 500 projection images spanning a 180 degree rotation. The DVC calculations were done using the AL-DVC software \cite{Yang2020aldvc} and they result in a displacement vector $\bm{u}$ for each point from which the 3D Green--Lagrange strain tensor
\begin{equation}
\epsilon_{ij} = \frac{1}{2} \left( \frac{\partial u_i}{\partial x_j} + \frac{\partial u_j}{\partial x_i} + \frac{\partial u_k}{\partial x_i} \frac{\partial u_k}{\partial x_j} \right)
\end{equation}
can be computed. From this tensor one can compute various strain invariants \cite{marsan2004scale}, such as the volumetric strain $\epsilon_{vol}$ and the shear invariant (second invariant of the deviatoric strain tensor) $\epsilon_{shear}$. The results we show here take both of these deformation modes into account by computing the total strain invariant $\epsilon_{tot} = \sqrt{\epsilon_{vol}^2 + \epsilon_{shear}^2}$. Additionally the local total strain rate $\dot{\epsilon}_{tot}$ is computed from the total strain at successive timesteps using numerical differentiation.

\section{Results}

Due to the high computation time of DVC calculations, 
three experiments were picked from a larger dataset for a full DVC analysis.
The stress-strain curves obtained from the experiments can be seen in Fig.~\ref{fig:setup}b and colored points in the plots correspond to single reconstructed tomography images.
The curves clearly show the three distinctive regimes: initial elastic behavior, the compaction with clear intermittent spikes in the strain rate (corresponding to avalanche-like behavior \cite{makinen2015avalanches}),  and the final densification.
In what follows we focus on one representative sample (similar data for other samples can be seen in the supplementary \cite{supplementary}).\\

The time-evolution of the local total strain $\epsilon_{tot}$ can be seen in Fig.~\ref{fig:imgSet}. One can clearly see the structure-dependent behavior of wood -- the wider softwood layers get compressed much more while the narrower hardwood layers maintain their shape.
The damage starts from hotspots in the softwood layers and then spreads throughout the layer as well as the other softwood layers. The mechanism does not seem to be damage propagation, rather new hotspots pop up at different parts of the layers.\\

While the strain localization in planes that correspond to the orientation of the annual rings is clearly visible to the naked eye, this can be seen even more clearly by computing the spatial autocorrelation function of the local total strain
\begin{equation} \label{eq:autocorrelation}
    C(\bm{s}) = \left\langle \epsilon_{tot}(\bm{r}) \epsilon_{tot}(\bm{r}+\bm{s}) \right\rangle_{\bm{r}}
\end{equation}
which can be seen in Fig.~\ref{fig:projection}a. To observe the time-evolution of the localization we fit a plane that corresponds to this orientation of high local strains (the blue plane in Fig.~\ref{fig:projection}a) and do a dimensionality reduction by projecting the voxel position vectors $\bm{r}$ to the line given by the unit normal of the plane $\bm{\hat{n}}$. Taking the average of the local total strain $\langle \epsilon_{tot} \rangle$ in 100 bins along this single dimension shows the localization evolution in the annual rings in a spacetime plot (top of Fig.~\ref{fig:projection}b).

While the evolution of local strain starting from the softwood parts of the annual rings closer to the top of the sample can already be seen from the previous figure, it can be further quantified by computing a localization measure. Here we use the the normalized standard deviation of this reduced quantity $\langle \epsilon_{tot} \rangle$ and define the localization measure as
\begin{equation}
    \eta = \frac{\Delta \langle \epsilon_{tot} \rangle}{\left\langle \langle \epsilon_{tot} \rangle \right\rangle}
\end{equation}
where $\Delta \langle \epsilon_{tot} \rangle$ denotes the standard deviation of the average local total strain over all the bins along the dimension given by $\bm{r} \cdot \bm{\hat{n}}$ and $\left\langle \langle \epsilon_{tot} \rangle \right\rangle$ denotes the average of the average total strain over all the bins. The evolution of this localization parameter can be seen in Fig.~\ref{fig:projection}c and it shows two things: firstly after an initial regime of localization due to damage nucleation the value of the localization parameter decreases, signifying strain delocalization. Secondly, the spreading of damage from one annual ring to another can be seen as a dip in the localization parameter. It is noteworthy that the changes in the parameter are fairly small.\\

\begin{figure}[t!]
    \centering
    \includegraphics[width=\columnwidth]{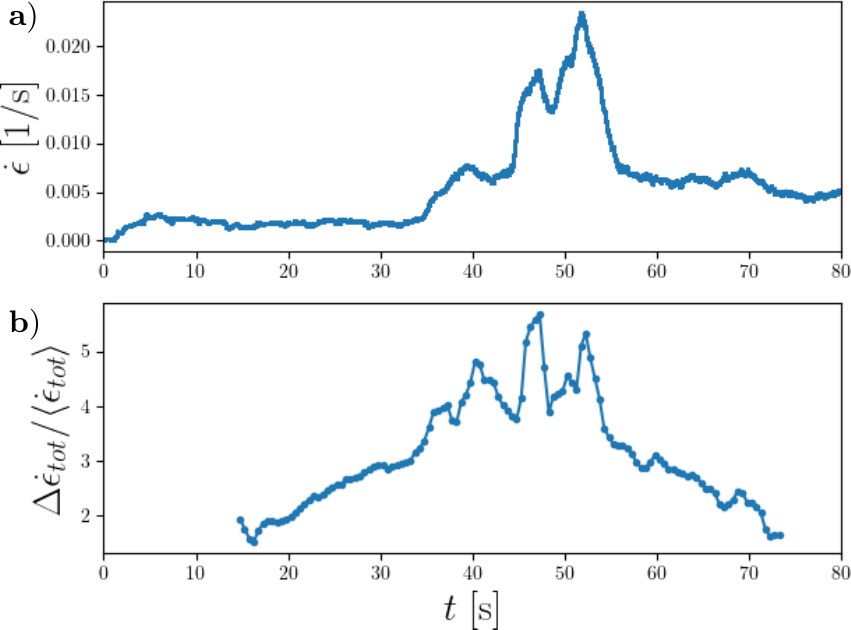}
    \caption{\textbf{a}) The global strain rate $\dot{\epsilon}$ as a function of time, showing the large variations (spikes) corresponding to the avalanches in the compaction regime. \textbf{b}) The standard deviation of the local total strain rate normalized by its average as a function of time, showing large local variations.}
    \label{fig:localization}
\end{figure}

The avalanche-like behavior in the compaction regime (second regime, flat part of the stress-strain curves) can be seen as spikes in the global strain rate (see Fig.~\ref{fig:localization}a). The local total strain rate distribution in the sample is fairly wide (fat-tailed) and this can be quantified by computing the normalized standard deviation of the local total strain rate $\Delta \dot{\epsilon}_{tot} / \langle \dot{\epsilon}_{tot} \rangle$. Plotting this next to the global strain rate (see Fig.~\ref{fig:localization}) shows that their time evolutions correlate very strongly, indicating that the avalanches are accompanied by large local variations in the local strain rate.
By computing the 1~\% and 10~\% fractiles of the local total strain we have verified that the general strain behavior can be seen to correspond fairly well to the behavior of the 1~\% fractile and only to a lesser extent to the behavior of the 10~\% fractile.\\

\begin{figure}
    \centering
    \includegraphics[width=\columnwidth]{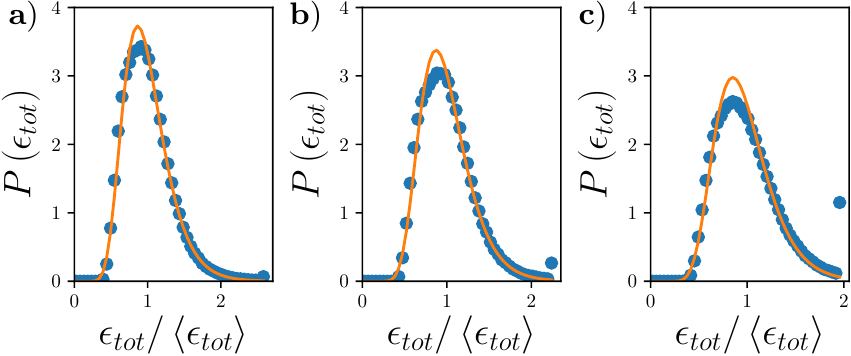}
    \caption{The probability density functions of the local total strain (blue dots) at three points in time (the panels corresponding to \textbf{a}) $t=20$ s, \textbf{b}) $t = 45$ s, and \textbf{c}) $t = 73$ s) as a function of the normalized total strain $\epsilon_{tot}/\langle \epsilon_{tot} \rangle$ and a log-normal fit (orange line).}
    \label{fig:distributions}
\end{figure}

Finally we compute the probability density functions of local strain at each timestep (see Fig.~\ref{fig:distributions}). The shape of the distributions seems close to a log-normal distribution except for a large spike at high strains at later stages of the experiments. The spike corresponds to regions of very high strains where the accuracy of the DVC algorithm gets reduced, resulting in strains of unity. Discarding this final spike and computing a maximum likelihood fit (orange lines) indeed shows that the log-normal distribution fits the data well. 

\section{Conclusions}

We have studied the time evolution of local strain in a wood sample under compression using fast 4D in-situ tomography and DVC. After an initial elastic regime we observe a compaction regime where the deformation is intermittent (strain rate spikes) and starts from "soft spots" in the softwood layers of the annual rings.
These spikes have previously \cite{makinen2015avalanches} been shown using AE to correspond to avalanche-like behavior with scalefree size statistics.
We note that the samples used here are several millimeters smaller than the ones in Ref.~\cite{makinen2015avalanches}.\\

Although most of the deformation is observed localized into the softwood layers what we actually observe considering the whole 3D local strain field is a spreading of the strain to all the softwood layers, instead of strain localization. The avalanches correlate well with large variations in the local strain rate suggesting localized avalanches. We interpret this to mean that the localization is happening in regions on planes parallel to the annual ring orientation. Any kind of dimensionality reduction (to 2D or 1D) hides this and shows only the strain spreading into all the annual rings.
The localization behavior observed shows that a key feature in the mechanical behavior of wood is the complex interaction between the annual rings. The localization in these is clearly not trivial, and in our small set of small samples we do not get a comprehensive idea apart from the fact that this interaction does exist and it is important.    \\

The log-normal form observed for the local strain distributions supports this interpretation as this distribution can result as a product of many random variables. Indeed log-normal strain distributions have been observed in metals \cite{tang2020lognormal, chen2021local} as a "universal" distribution due to (plastic) strain accumulation. Also in sea ice deformation, which shows similar fat-tailed local strain rate distributions, the multifractal behavior \cite{stanley1988multifractal} of the strain rate fields has been mapped to a log-normal multiplicative cascade model \cite{marsan2004scale}. This log-normal observation leaves however open the issue as to what happens to the local strain distribution close to failure as the increments become more correlated.  

This view is also compatible with the previously observed
robust power-law distribution for the AE energies irrespective of the event rate.
The structure-dependent softwood layer collapses seem to manifest themselves only in the variations of the AE event rate.\\

Our results show that when considering inherently bulk phenomena (not just surface effects), such as avalanche behavior, the use of 3D imaging can be extremely useful in unveiling the true nature of the material behavior.
An increase in the spatial and temporal resolution would still be needed for the direct observation of the avalanche phenomena in the sample. Likewise, the sample size could be increased well over the current one here. 

\begin{acknowledgments}
We acknowledge the European Synchrotron Radiation Facility for provision of synchrotron radiation facilities and we would like to thank Marco di Michiel for assistance in using beamline ID15A.
We thank Simo Huotari and Heikki Suhonen for their help in the application and data analysis process.

M.J.A. acknowledges support from the Academy of Finland (Center of Excellence program, 278367 and 317464). 
M.J.A and T.M. acknowledge funding from the European Union Horizon 2020 research and innovation programme under Grant Agreement No. 857470 and from European Regional Development Fund via Foundation for Polish Science International Research Agenda PLUS programme Grant No. MAB PLUS/2018/8.
T.M. also acknowledges funding from The Finnish Foundation for Technology Promotion.
J.K. acknowledges the funding from Academy of Finland (308235) and Business
Finland (211715).
Aalto Science IT project is acknowledged for computational resources.
\end{acknowledgments}

%\bibliography{refs}% Produces the bibliography via BibTeX.

\end{document}

\preprint{APS/123-QED}

\title{Supplementary material: Wood compression in 4D in-situ tomography}% Force line breaks with \\

\author{Tero M\"{a}kinen}
 \email{Corresponding author\\tero.j.makinen@aalto.fi}
\affiliation{%
 Department of Applied Physics, Aalto University,
 P.O.  Box 11100, 00076 Aalto, Espoo, Finland
}%
\affiliation{NOMATEN Centre of Excellence, National Centre for Nuclear Research, ul. A. Soltana 7, 05-400  Otwock-\'{S}wierk, Poland}
\author{Alisa Halonen}
\affiliation{%
 Department of Applied Physics, Aalto University,
 P.O.  Box 11100, 00076 Aalto, Espoo, Finland
}%
\author{Juha Koivisto}%
\affiliation{%
 Department of Applied Physics, Aalto University,
 P.O.  Box 11100, 00076 Aalto, Espoo, Finland
}%

\author{Mikko J. Alava}%
\affiliation{%
 Department of Applied Physics, Aalto University,
 P.O.  Box 11100, 00076 Aalto, Espoo, Finland
}%
\affiliation{NOMATEN Centre of Excellence, National Centre for Nuclear Research, ul. A. Soltana 7, 05-400  Otwock-\'{S}wierk, Poland}

\date{\today}% It is always \today, today,

\maketitle

%\tableofcontents

\section{Additional experiments}

In addition to the one experiment discussed in detail in the main text we show here the results from two additional experiments. We denote the sample in the main text as sample A, and the additional ones shown in this supplementary as sample B and C.

The reconstructed volumes of samples B and C are shown in Fig.~\ref{fig:samples} and the whole timeseries of reconstructed volumes for each experiment is shown in supplementary videos (filenames \texttt{reconstruction\_X.mp4} where \texttt{X} denotes the sample name).

Similarly the full timeseries of local total strains is shown in supplementary videos (filenames \texttt{strain\_X.mp4} where \texttt{X} denotes the sample name).

Additionally the Figs.~3 and 4 of the main paper are recreated here for sample~B (Fig.~\ref{fig:006}) and sample~C (Fig.\ref{fig:065}).

\begin{figure}[h!]
    \centering
    \includegraphics{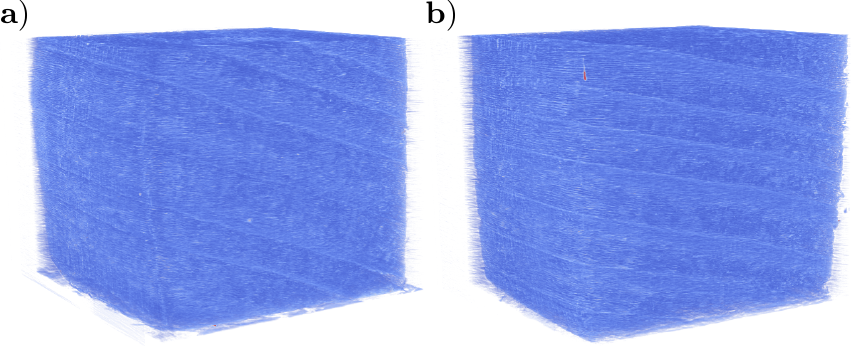}
    \caption{\textbf{a}) The reconstructed volume of sample B showing an annual ring structure similar to sample A.
    \textbf{b}) The reconstructed volume of sample C showing slimmer annual rings and a more perpendicular orientation of the rings in relation to the compression axis.}
    \label{fig:samples}
\end{figure}

\section{Annual rings}

While the annual ring structure in samples A and B is fairly similar (similar number of rings and similar orientation), sample C differs from those. The annual rings are slimmer and therefore there are more of them in the sample, and additionally they are oriented more perpendicular to the compression direction.\\

\begin{figure*}[p!]
    \centering
    \includegraphics[width=\textwidth]{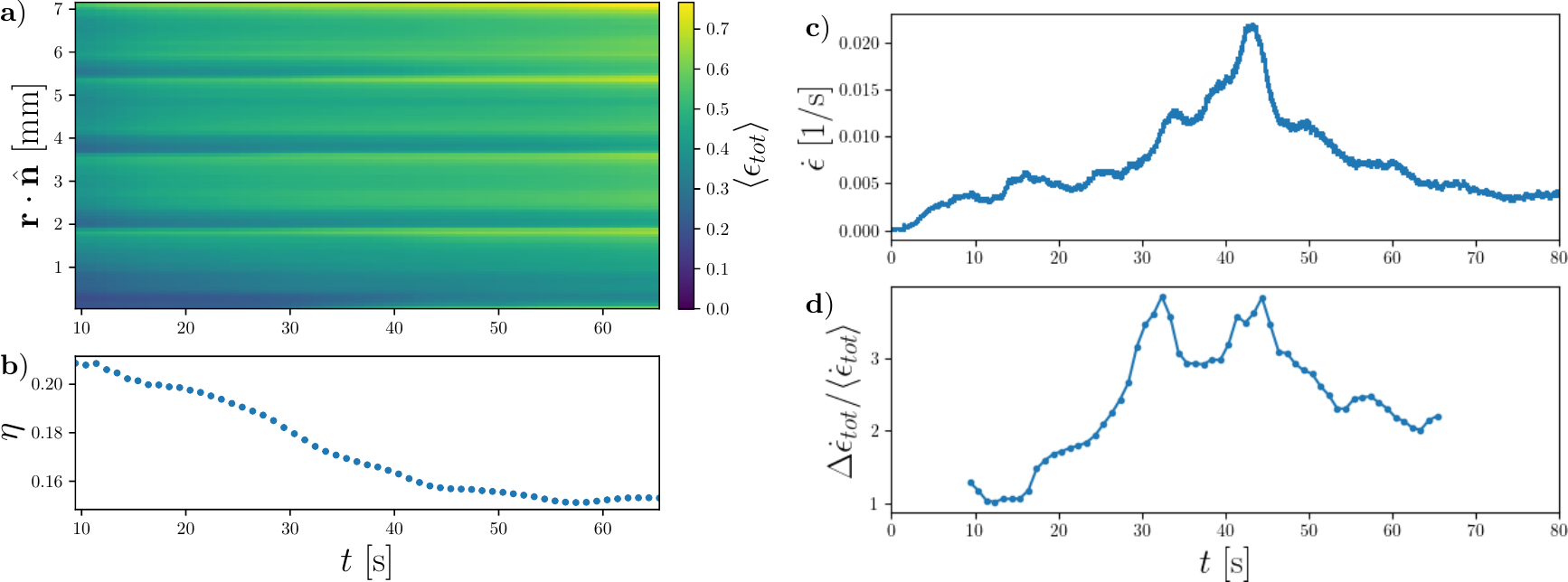}
    \caption{Sample B:
    \textbf{a}) A spacetime plot of the average local strain $\left\langle \epsilon_{tot} \right\rangle$ in the direction perpendicular to the fitted plane (same procedure as in Fig.~3a of the main text), described by the projection of the position vector $\bm{r}$ onto the unit normal of the plane $\hat{\bm{n}}$.
    \textbf{b}) The localization measure $\eta$ as a function of time $t$ (time axis aligned with panel a) showing the delocalization of strain, similarly to the Fig.~3c of the main text.
    \textbf{c}) The global strain $\dot{\epsilon}$ as a function of time, showing the large variations (spikes) corresponding to the avalanches in the compaction regime.
    \textbf{d}) The standard deviation of the local total strain rate normalized by its average as a function of time, showing large local variations similarly to the Fig.~4b of the main text.}
    \label{fig:006}
\end{figure*}

Looking at Fig.~\ref{fig:006} we can see that sample B rougly follows the behavior of sample A presented in the main text. The spreading of the strain into the annual rings occurs in a slightly smoother fashion (see Fig.~\ref{fig:006}a) but the localization measure (Fig.~\ref{fig:006}b) still shows minor downward jumps, similar to Fig.~3a in the main text. Similarly the global strain rate (Fig.~\ref{fig:006}c) shows slightly smoother behavior, but the spikes still correlate well with the variations in the local strain rate (Fig.~\ref{fig:006}d).\\

\begin{figure*}[p!]
    \centering
    \includegraphics[width=\textwidth]{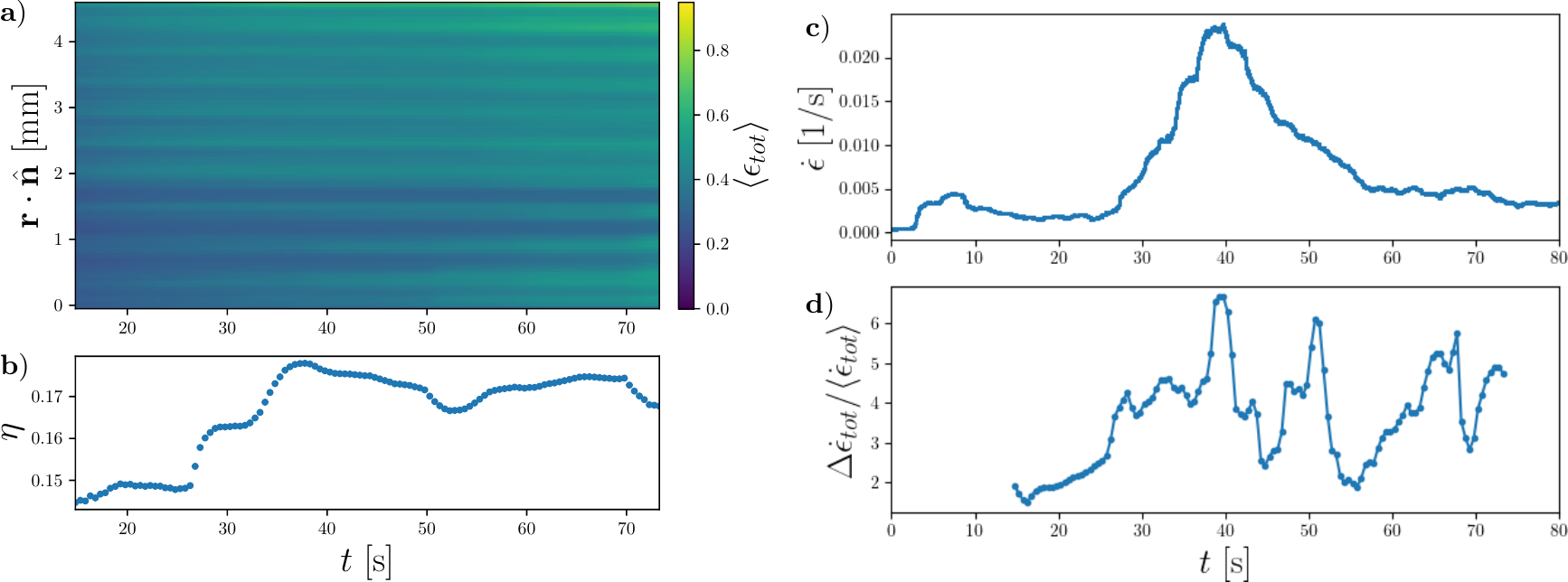}
    \caption{Sample C:
    \textbf{a}) A spacetime plot of the average local strain $\left\langle \epsilon_{tot} \right\rangle$ in the direction perpendicular to the fitted plane (same procedure as in Fig.~3a of the main text), described by the projection of the position vector $\bm{r}$ onto the unit normal of the plane $\hat{\bm{n}}$.
    \textbf{b}) The localization measure $\eta$ as a function of time $t$ (time axis aligned with panel a) showing a mixture of localization and delocalization of strain, slightly differing from the picture of Fig.~3c of the main text.
    \textbf{c}) The global strain $\dot{\epsilon}$ as a function of time, showing less spiked profile compared to Fig.~4a of the main text of Fig.~\ref{fig:006}c.
    \textbf{d}) The standard deviation of the local total strain rate normalized by its average as a function of time, showing large local variations.}
    \label{fig:065}
\end{figure*}

In sample C the annual ring orientation slightly suppresses the shear deformation and the slimmer rings (Fig.~\ref{fig:065}a) have a clear effect on the localization (Fig.~\ref{fig:065}b). Before the global strain rate maximum (at around $t = 40$ s) the localization measure shows an increase and after that it is roughly constant with some minor downward jumps.
The global strain rate behavior is much smoother than with samples A and B but the maximal variations in the local strain rate still correlate with the global strain rate spike. However after the global strain rate maxima there still are large local strain rate fluctuations that do not correlate with the global strain rate behavior.\\

We interpret this to mean that the interplay of a larger number of annual rings smoothens out the localized jerky behavior.


%\bibliography{refs}% Produces the bibliography via BibTeX.